\documentclass[twocolumn,nofootinbib,prd,aps,tightenlines]{revtex4}

\usepackage{graphicx}
\usepackage{pstricks}
\usepackage{bm}

\def\dfrac#1#2{ \frac{#1}{#2} }
\def\ket#1{\left| #1\right\rangle}

{\count255=\time\divide\count255 by 60 \xdef\hourmin{\number\count255}
 \multiply\count255 by-60\advance\count255 by\time
 \xdef\hourmin{\hourmin:\ifnum\count255<10 0\fi\the\count255}}

\newcommand{\nn}{\nonumber \\ }

\begin{document}

\title{Dispersion Relation Bounds for $\pi\,\pi$ Scattering}

\author{Aneesh V.~Manohar}
\affiliation{Department of Physics, University of California at San Diego,
 La Jolla, CA 92093}

\author{Vicent Mateu}
\affiliation{Departament de F\'\i sica Te\`orica, IFIC, Universitat
de Val\`encia-CSIC, Apt. Correus 22085, E-46071 Val\`encia, Spain}

\date{\today\quad\hourmin}

\begin{abstract}
\noindent Axiomatic principles such as analyticity, unitarity, and crossing symmetry constrain the second derivative of the  $\pi\,\pi$ scattering amplitudes in some channels to be positive in a region of the Mandelstam plane. Since this region lies in the domain of validity of chiral perturbation theory, we can use these positivity conditions to bound linear combinations of $\bar{l}_{1}$ and $\bar{l}_{2}$. We compare our predictions with those derived previously in the literature using similar methods. We compute the one-loop $\pi\,\pi$ scattering amplitude in the linear sigma model (LSM) using the $\mathrm{\overline{MS}}$ scheme, a result hitherto absent in the literature. The  LSM values for $\bar{l}_{1}$ and $\bar{l}_{2}$ violate the bounds for small values of $m_\sigma/m_\pi$. We show how this can occur, while still being consistent with the axiomatic principles.
\end{abstract}

\maketitle

\section{Introduction}
\label{sec:Introduction}

Quantum chromodynamics (QCD) with two light quark  flavors has an approximate $SU(2)_L \times SU(2)_R$ chiral symmetry which is spontaneously broken to its diagonal vector subgroup $SU(2)_V$, leading to an isotriplet of pseudo-Goldstone bosons, the pions.  Low-energy pion dynamics, particularly elastic pion-pion scattering, encodes useful information about the confining dynamics of the strong interactions.

The standard technique to study pion dynamics at very low energies with effective field theories was proposed in Ref.~\cite{Wein} (see also Ref.~\cite{ccwz}) and systematized as an expansion in powers of momentum and quark masses in Ref.~\cite{Gasser1}. This effective theory is known as chiral perturbation theory ($\chi$PT). It is formulated in terms of a Lagrangian whose only degrees of freedom are pions and which incorporates the symmetries of QCD, including spontaneously broken chiral symmetry~\cite{ccwz}.

At lowest order in the chiral expansion, the physical observables are determined in terms of two parameters, the pion decay constant and the pion mass. If one goes beyond the lowest order, a number of low-energy constants (LECs) $l_{i}$ not fixed by symmetries must be included. These can be determined by fitting to experimental data (for the best determination, see Ref.~\cite{Colangelo}) or estimated by vector-meson dominance \cite{RChT,Pennington:1994kc}, but both methods have large uncertainties.

An alternative formulation of $\pi\,\pi$ scattering can be obtained based only on axiomatic principles of quantum field theory, such as analyticity, unitarity, and crossing symmetry. This allows one to obtain relations between observable quantities that must hold, regardless of the theory used for the description of the phenomenon under study. Of course, one of the usual benefits of an effective theory approach is that many of these principles are automatically satisfied by the scattering amplitudes computed using the effective theory. Nevertheless, there is still useful information missing in the effective theory approach, and one obtains interesting results by studying the constraints imposed by axiomatic principles on the effective Lagrangian. Analyticity and unitarity can be exploited to write the well known dispersion relations for the scattering amplitudes. These, together with crossing symmetry, can be converted into positivity conditions on scattering amplitudes, which in turn can be combined with the $\chi$PT predictions to give bounds on the $\bar{l}_{1}$ and $\bar{l}_{2}$ LECs in the chiral Lagrangian at order $p^{4}$.

Two-flavor $\chi$PT was combined with axiomatic principles in Ref~\cite{Ananthanarayan:1994hf}, which analyzed constraints on $s$ and $p$ partial-wave amplitudes in the framework of dispersion relations. The analysis was done  in $\chi$PT  at the one-loop level. In Ref.~\cite{Dita:1998mh} this study was extended to cover all three isospin amplitudes of $\pi\,\pi$ scattering at the  two-loop level in $\chi$PT. The best bounds were found for positivity conditions on full amplitudes (in contrast with partial-wave amplitudes), and we follow this approach in the present work. However, we find inconsistencies in the domain of applicability of the positivity constraints used in Ref.~\cite{Dita:1998mh} which will be explained in Sec.~\ref{sec:Comparison}. Similar bounds were first found in Ref.~\cite{Pennington:1994kc} in the context of the  Froissart-Gribov representation for the scattering lengths. More recently, in Ref.~\cite{Distler:2006if}, the very same bounds of Ref.~\cite{Pennington:1994kc} were  rediscovered using the same procedure as in Ref.~\cite{Dita:1998mh} but using a more restricted domain of validity (in the Mandelstam plane)  of the positivity constraint.  References~\cite{Pennington:1994kc,Distler:2006if} both used one-loop $\chi$PT amplitudes. We show that the methods of Ref.~\cite{Pennington:1994kc} and Refs.~\cite{Dita:1998mh,Distler:2006if} are equivalent, and we improve the bounds by properly using the domain of validity considered in Ref.~\cite{Dita:1998mh}, which is bigger than that considered in Ref.~\cite{Distler:2006if}.

In Ref.~\cite{Comellas:1995hq} a different approach was followed for putting bounds on some $\chi$PT parameters. QCD
inequalities on Green functions of quark bilinear currents were used to obtain relations (inequalities) that 
involve light quark masses, the quark condensate, and some LECs. With our method we are insensitive to the quark mass and condensate, since these are lowest order quantities, and our analysis starts at $\mathcal{O}(p^4)$. On the other hand, since our study relies on scattering amplitudes, we only make use of the chiral Lagrangian when vector, axial-vector, and scalar sources are switched off (one always needs the scalar source for giving masses to the pions). In fact we can only give bounds for the $\mathcal{O}(p^4)$ LECs of operators containing only pion fields, $\bar{l}_{1}$ and $\bar{l}_{2}$, and so our results do not overlap with theirs. 

One of the most popular models used in the literature for the study of pion dynamics is the linear sigma model (LSM), introduced in the sixties by Gell-Mann and Levy~\cite{LSM}. In this model, spontaneous chiral symmetry breaking is driven by a scalar particle $\sigma$ acquiring a nonvanishing vacuum expectation value. The LSM Lagrangian is renormalizable and thus has a reduced (finite) number of parameters compared with the most general chiral Lagrangian. It shares the same symmetries as $\chi$PT but has an additional ($\sigma$) particle in its spectrum. If the $\sigma$ mass is sufficiently greater than that of the pions, it can be formally integrated out of the action, leaving behind the $\chi$PT Lagrangian, with all the low-energy constants having specific values which can be predicted in terms of the finite number of parameters of the LSM.

The values for $\bar{l}_{1}$ and $\bar{l}_{2}$ predicted by the LSM do not satisfy the dispersion relation bounds for  low values of the $\sigma$ mass. We will demonstrate that the LSM is perfectly consistent with the dispersion relation bounds and that the apparent contradiction results because for low values of the $\sigma$ mass, integrating out the $\sigma$ is not valid, or equivalently, that higher order terms in the chiral expansion cannot be neglected.

This paper is organized as follows: in Sec.~\ref{sec:Dispersion} we describe the general features of $\pi\,\pi$ scattering such as crossing and analyticity, and derive the corresponding dispersion relations; in Sec.~\ref{sec:bounds} we show how dispersive integrals imply a positivity condition for the second derivative of the scattering amplitude; in Sec.~\ref{sec:ChiPT} we convert those positivity conditions into bounds for the chiral LECs $\bar{l}_{1,2}$; in Sec.~\ref{sec:Comparison} we compare our results with previous analyses, and in Sec.~\ref{sec:sigma-model} we resolve the apparent contradiction between the LSM prediction for the chiral LECs and the bounds previously found; our conclusions are summarized in Sec.~\ref{sec:Conclusions}. In Appendix~A we show the relation between the methods of Refs.~\cite{Pennington:1994kc} and \cite{Distler:2006if}; in Appendix~B we calculate the one-loop $\pi\,\pi$ scattering amplitude in the LSM renormalized in the $\overline{\mathrm{MS}}$ scheme.

\section{Dispersion relations for $\pi\,\pi$ scattering}
\label{sec:Dispersion}

In this section, we find the region of the Mandelstam $s-t$ plane in which the $\pi\,\pi$ scattering amplitude is analytic, and derive the corresponding dispersion relations. In a scattering process such as $a(p_{a})+b(p_{p})\to c(p_{c})+d(p_{d})$ the (nonindependent) Mandelstam variables are defined as
\begin{eqnarray}
&&s \,=\, (p_{a}+p_{b})^{2}\,,\quad t\,=\,(p_{a}-p_{c})^{2}\,,\quad 
u\,=\,(p_{a}-p_{p})^{2} \nn
&& s+t+u\,=\, \sum_{i=1}^{4}m_{i}^{2}\,.
\end{eqnarray}
Reversing the order of the two final (or initial) states amounts to exchanging $t$ and $u$.

We begin by briefly reviewing a few properties of $\pi\,\pi$ scattering. For further details the reader is referred, for instance, to Ref.~\cite{Peterson}. The three pionic states can be labeled either by $I_{3}=-1,\,0,\,1$ or by Cartesian indices $a=1,\,2,\,3$. Both sets of states are linearly related between them and to the physical pion states:
\begin{eqnarray}
\ket{\pi^{\pm}} \,=\,\frac{1}{\sqrt{2}}\left(\ket{\pi^{1}} \mp \ket{\pi^{2}} \right)\,, &  & 
\ket{\pi^{0}} \,=\,\ket{\pi^{3}} \,,\nonumber \\
\ket{1,\pm1} \,=\,\mp\ket{\pi^{\pm}} \,, &  & 
\ket{1,0} \,=\,\ket{\pi^{0}} \,,
\end{eqnarray}
where $\ket{\pi^a}$ denotes the Cartesian basis, and $\ket{1,I_3}$ denotes the isospin basis states. Isospin invariance implies that there are only three linearly independent scattering amplitudes in the $I=0,1,2$ channels, and crossing symmetry relates them to each other, so they can all be described by a single function of $s$ and $t$. In the Cartesian basis we can write the Chew-Mandelstam formula
\begin{eqnarray}
T(a\, b\to c\, d)&=& A(s,t,u)\,\delta^{ab}\delta^{cd}+A(t,s,u)\,\delta^{ac}\delta^{bd}\nn
&&+\,A(u,t,s)\, \delta^{ad}\delta^{bc}\,,\label{eq:chew-man}
\end{eqnarray}
where crossing symmetry implies $A(x,y,z)=A(x,z,y)\equiv A(x,y)=A(x,4\,m^{2}-x-y)$ where $m$ is the pion mass. The function $A$ is related to the isospin amplitudes through
\begin{eqnarray}
T^{0}(s,t)&=&3\, A(s,t)+A(t,s)+A(u,s)\,,\nn
T^{1}(s,t)&=& A(t,s) - A(u,s)\,,\nn
T^{2}(s,t)&=& A(t,s) + A(u,s)\,.
\label{eq:isospin}
\end{eqnarray}
The $I=0,2$ amplitudes are symmetric under the exchange of the final states, whereas the $I=1$ is antisymmetric: $T^{(0,2)}(s,t)=T^{(0,2)}(s,u)$, $T^{1}(s,t)=-\,T^{1}(s,u)$. 

The isospin amplitudes in the different kinematic channels are also linearly related. For our present purposes, we only need the relation with the crossed $u$-channel. This follows directly from Eq.~(\ref{eq:isospin}) and can be conveniently displayed in matrix notation \cite{Roy:1971tc}
\begin{eqnarray}
T^{\, I}(s,t)&=& C_{u}^{II^{\prime}}T^{\, I^{\prime}}(u,t)\,,\nn 
C_{u}^{II^{\prime}}C_{u}^{I^{\prime}J}&=&\delta_{IJ}\,,\nn 
C_{u}&=&\frac{1}{6}\left(\!\begin{array}{ccc}
\phantom{\,+}2 & -\,6 & 10\\
-\,2 & \phantom{+\,}3 & 5\\
\phantom{+\,}2 & \phantom{+\,}3 & 1\end{array}\right)\,,\label{eq:crossing-u}
\end{eqnarray}
where, as expected, the crossing-matrix $C_{u}$ is its own inverse. $T^{\, I}(s,t)$ is the scattering amplitude with isospin $I$ in the $s$-channel, and $T^{\, I^{\prime}}(u,t)$ is the amplitude with isospin $I^\prime$ in the $u$-channel.

Axiomatic principles can be used to show that scattering amplitudes are analytic in the full complex $s$ plane except for possible isolated points, due to single-particle exchange, and branch cuts, due to unitarity. For our purposes we only need to know the position $s_{0}$ of the first branch point along the real axis of the complex $s$ plane. There is then a branch  cut along the real $s$-axis for $s \ge s_0$. Any other singularities along the real $s$-axis will be along this cut.  The remaining branch cuts will be determined by crossing symmetry.

\begin{figure}
\begin{center}
\includegraphics[width=7.9cm]{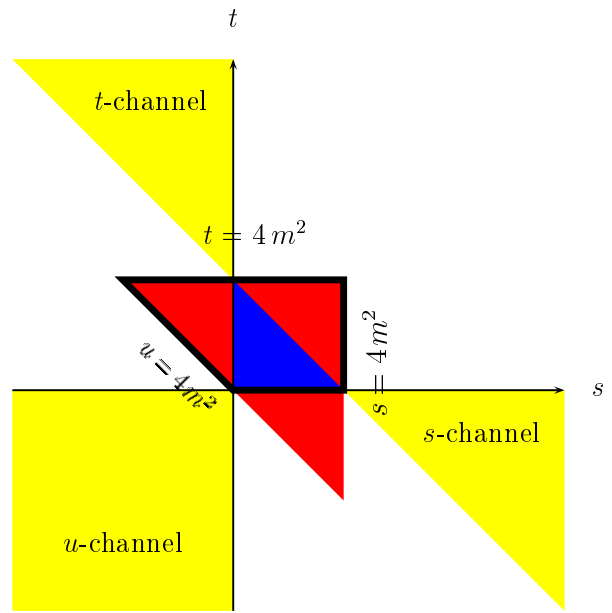}
\end{center}
\caption{Mandelstam plane for $\pi\,\pi$ scattering. The small (blue) triangle in the center is the Mandelstam triangle. The big triangle (red and blue area) is the region free from singularities. The outer regions (yellow) denote the physical regions for the three crossed channels. The region bounded by the thick black line corresponds to the area $\mathcal{A}$ in which the positivity conditions are satisfied.\label{fig:pipi-diagram} }
\end{figure}

Let us concentrate on the $s$-channel keeping $t$ fixed. Unitarity ensures that for real $s$, the scattering amplitude only develops an imaginary part above the lowest mass threshold of possible intermediate states.\footnote{Above threshold, the physical scattering amplitude is defined as the value given by approaching the cut from above, $T^{\mathrm{phys}}(s,t)\,=\, T(s+i\,\epsilon,t)$, with $\epsilon \to 0$. This corresponds to the Feynman $i\,\epsilon$ prescription for propagators.} In our case the threshold corresponds to two-pion states, \emph{i.e.} $s_{0}=4\, m^{2}$. This means that above the production threshold (for physical amplitudes) the scattering amplitude is complex. Since below threshold, the amplitude is real and analytic away from the real axis, it follows from the Schwarz reflection principle that  $T^{*}\left(s+i\,\epsilon\right)=T\left(s-i\,\epsilon\right)$ and hence $T\left(s+i\,\epsilon\right)-T\left(s-i\,\epsilon\right)=2\,  i\,\mathrm{Im}\,T\left(s+i\,\epsilon\right)\neq0$. This means there must be a branch point at $s=4\, m^{2}$, and a discontinuity in the amplitude along the real  axis for $s > 4\,m^2$. We will choose the branch cut to run along the real $s$-axis, because as already explained, the other branch points due to higher mass thresholds (\emph{e.g.} four-pion state $s_{1}=16\, m^{2}$) or singularities due to single-particle states (e.g. $\rho$ exchange $s_{\rho}=m_{\rho}^{2}$) will lie along it. We conclude that our amplitude is nonanalytic for $s>4\, m^{2}$, regardless of the value of $t$. The amplitude must also reproduce the singularities in the crossed channels, so it is nonanalytic for $s,t,u>4\,m^2$. The region in the $s-t$ plane where the amplitude is analytic is limited to the inside of the  triangle defined by the conditions $s,\, t\le4\, m^{2}$, $s+t\ge0$. $4\, m^{2}$ is referred to as the normal threshold, associated to the production of two pions.  In Refs.~\cite{Ananthanarayan:1994hf,Distler:2006if} it is assumed that the amplitude is only analytic between the normal threshold and the abnormal threshold, corresponding to $s,\, t,\, u=0$. The region delimited by the condition $0<s,\, t,\, u<4\, m^{2}$ is known as the Mandelstam triangle (see Fig.~\ref{fig:pipi-diagram}). 

However it has been proved \cite{Eden-book} using very general arguments that rely on perturbation theory to all orders (i.e.\ that are true for every single Feynman diagram), that the amplitude becomes nonanalytic only above the normal threshold, and that nothing particular happens at $s=0$. The region bounded by $s,t,u < 4\,m^2$ is the larger triangle shown in Fig.~\ref{fig:pipi-diagram}. This is the main difference between our method and that of Ref.~\cite{Distler:2006if}. We use analyticity in a larger domain, and so obtain more restrictive conditions on the scattering amplitude. Reference~\cite{Dita:1998mh} uses the same analyticity domain as we do. However, in their numeric computations, they include points outside this region, which is not justified.

\begin{figure}
\begin{center}
\includegraphics[width=8cm]{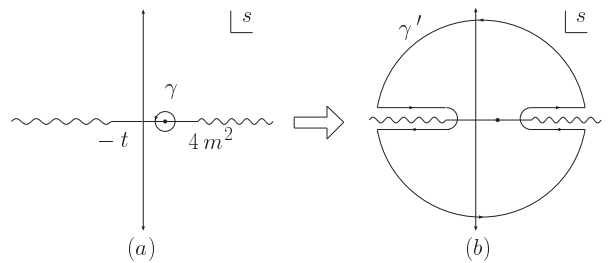}\end{center}
\caption{Contour integrals leading to the fixed-$t$ dispersion relations.
\label{fig:contour-pipi}}
\end{figure}

The derivation of the dispersion relation is quite straightforward and is very nicely explained, for instance, in Ref.~\cite{Martin-Spearman}. For our derivation we consider $t$ as a fixed parameter. We can then use Cauchy's theorem to write
\begin{equation}
T^{I}(s,t)\,=\,\frac{1}{2\pi\, i}\oint_{\gamma}\mathrm{d}x\,\frac{T^{I}(x,t)}{x-s}\,,
\label{eq:first-contour}
\end{equation}
wherever the amplitude is analytic in a neighborhood (in $s$) of the point $(s,t)$, and where the contour $\gamma$ encloses the point $x=s$ [\,see Fig.~\ref{fig:contour-pipi}(a)\,]. 
Then $t\le4\, m^{2}$, and if $s>0$, we have to use $s\to s+i\,\epsilon$, as already mentioned. From the results of Ref.~\cite{Eden-book} we infer that fixed-$t$ dispersion relations hold for $t<4\,m^2$, but using solely axiomatic principles it can be shown (Ref.~\cite{Martin:1965jj}) that they are at least valid in the interval $-\,28\,m^2\le t \le 4\,m^2$, which will be adequate for our purposes. For fixed $t$, we have along the real $s$-axis a right-hand branch cut for $s>4\, m^{2}$ and a left-hand branch cut for $s<-\, t$. The $\gamma$ contour in Eq.~(\ref{eq:first-contour}) can be deformed into $\gamma^{\,\prime}$, as shown in Fig.~\ref{fig:contour-pipi}(b) in order to express the integral in terms of the discontinuity of the amplitude along the real axis. In order to do this, the amplitude must fall sufficiently rapidly that the contribution from the contour at infinity vanishes. If it does not, we can perform $n$ derivatives (subtractions) to increase the convergence at infinity,
\begin{equation}
\frac{\mathrm{d}^{n}}{\mathrm{d}s^{n}}T^{I}(s,t)\,=\,\frac{n!}{2\pi i}
\oint_{\gamma}\mathrm{d}x\,\frac{T^{I}(x,t)}{(x-s)^{n+1}}\,.\label{eq:subtraction-scattering}
\end{equation}
For large enough $n$ that the contour at infinity does not contribute, one finds after some straightforward manipulations that
\begin{eqnarray}
&&\frac{\mathrm{d}^{n}}{\mathrm{d}s^{n}}T^{I}(s,t)=\frac{n!}{\pi}\int_{4\,m^{2}}^{\infty}
\mathrm{d}x\Biggl[\,\frac{\delta^{II'}}{(x-s)^{n+1}}\nn
&&+(-1)^{n}\,\frac{C_{u}^{II'}}{(x-u)^{n+1}}\Biggr]
\mathrm{Im}\, T^{I'}(x+i\,\epsilon,t)\,.\label{eq:disp_final}
\end{eqnarray}
The first term is from the discontinuity across the right-hand cut. The second term is from the discontinuity across the left-hand cut, rewritten using crossing symmetry and Eq.~(\ref{eq:crossing-u}) to relate the $s$-channel discontinuity in the unphysical region $s<0$ to the $u$-channel discontinuity in the physical region.

The best constraint comes from using Eq.~(\ref{eq:disp_final}) with the smallest possible value of $n$. The Froissart bound \cite{Froissart} fixes the minimum number of subtractions needed for pion-pion scattering to $n=2$. Clearly, if we restrict ourselves to $s<4\, m^{2}$ and $s+t>0$, both denominators in Eq.~(\ref{eq:disp_final}) are positive, and if $n$ is an even number (for instance, in our case $n=2$) the relative sign is also positive, except for the sign of $C_{u}^{II^\prime}$.

\section{Bounds implied by the dispersion relation}
\label{sec:bounds}

Each isospin amplitude admits a partial-wave expansion. In the case of spin-zero particles, the amplitude depends only on the scattering angle $\theta$, defined as the angle between the three-momenta of the first initial and final pions, in the center of mass frame. Expanding in terms of Legendre polynomials $P_{\ell}$ we get: 
\begin{eqnarray}
T^{I}(s,t)&=&\sum_{\ell=0}^{\infty}\,(2\,\ell+1){\, f}_{\ell}^{I}(s)P_{\ell}(\cos\theta)\,\nn
&=&\sum_{\ell=0}^{\infty}\,(2\,\ell+1)\, f_{\ell}^{I}(s)P_{\ell}\!
\left(1+\frac{2\, t}{s-4\, m^{2}}\right)\,,\label{eq:partial_s}
\end{eqnarray}
where $f_{\ell}^{I}(s)$ denotes the partial-wave amplitudes. The optical theorem implies
\begin{equation}
\mathrm{Im}\, f_{\ell}^{I}(s)\,=\, s\,\beta(s)\,\sigma_{\ell}^{I}(s)\,\ge\,0\,,
\label{eq:forward}
\end{equation}
where $\beta(s)=\sqrt{1-\frac{4\,m^{2}}{s}}$ is the velocity of the pions in the center of mass frame,  and $\sigma_\ell^I$ are the partial-wave cross-sections in a given isospin channel. Equation~(\ref{eq:forward}) gives
\begin{equation}
\mathrm{Im}\, T^{I}(s,t)\,=\sum_{\ell=0}^{\infty}\,(2\,\ell+1)s\,\beta(s)\,
\sigma_{\ell}^{I}(s)P_{\ell}\!\left(1+\frac{2\, t}{s-4\, m^{2}}\right)\,.\label{eq:imF}
\end{equation}
The partial-wave expansion of the absorptive part converges in the large Lehmann-Martin ellipse, which, when projected onto real $s$ translates into the interval $-\,4\,m^2<s<60\,m^2$.  We also need to make sure the absorptive part is positive. In Eq.~(\ref{eq:disp_final}), the region of integration is $s>4\, m^{2}$, and as pointed out in Ref.~\cite{Ananthanarayan:1994hf}, since $P_{\ell}(z)>1$ for $z>1$ for all $\ell$, if we restrict ourselves to $t>0$, each partial-wave contribution to the imaginary part is positive and so the full imaginary part is itself positive. As noted in Ref.~\cite{Distler:2006if}, one can find certain linear combinations $\sum a_{I}\,T^{I}$ with $a_{I}\ge0$, $\sum a_{I}\,C_u^{IJ}\,T_J\equiv\sum_J b_J \,T_J$ with $b_J=\sum_I  a_{I}\, C_{u}^{IJ} \ge0$. For these linear combinations, the two terms in brackets in Eq.~(\ref{eq:disp_final}) give a positive contribution. Hence, for these linear combinations, in the region $\mathcal{A}$ defined as $s,\, t<4\,m^{2}$, $t>0$, and $s+t>0$ (see Fig.~\ref{fig:pipi-diagram}) the right-hand side of Eq.~(\ref{eq:disp_final}) for $n=2$ is also positive. 

There are three linear combinations which satisfy the positivity condition,  corresponding to the physical processes $\pi^0 \pi^0 \to \pi^0 \pi^0$, $\pi^+ \pi^+ \to \pi^+ \pi^+$, and $\pi^+ \pi^0 \to \pi^+ \pi^0$. These results are in fact expected and can be deduced without any mention of isospin amplitudes. The optical theorem ensures that for processes with the same initial and final particles $a+b\to a+b$, the imaginary part of each partial-wave is positive definite. The crossed $u$-channel for those processes has equal initial and final states as well, $a+\bar{b}\to a+\bar{b}$, so for such processes, the imaginary part along the right- and left-hand cuts will be always positive. The positivity conditions for the three processes are
\begin{eqnarray}
0 &\le&  \frac{\mathrm{d}^{2}}{\mathrm{d}s^{2}}T\left(\pi^{0}\pi^{0}\to
\pi^{0}\pi^{0}\right)[\,(s,t)\in\mathcal{A}\,]\,, \nn
0 & \le&  \frac{\mathrm{d}^{2}}{\mathrm{d}s^{2}}T\left(\pi^{+}\pi^{0}\to
\pi^{+}\pi^{0}\right)[\,(s,t)\in\mathcal{A}\,]\,,\nn
0 & \le&  \frac{\mathrm{d}^{2}}{\mathrm{d}s^{2}}T\left(\pi^{+}\pi^{+}\to
\pi^{+}\pi^{+}\right)[\,(s,t)\in\mathcal{A}\,]\,,\label{eq:positive}
\end{eqnarray}
corresponding to $\frac{2}{3}\,T^{(2)}(s,t)+\frac{1}{3}\,T^{(0)}(s,t)$,
$\frac{1}{2}\,T^{(2)}(s,t)+\frac{1}{2}\,T^{(1)}(s,t)$, and $T^{(2)}(s,t)$,
respectively.

\section{Bounds for $\mathbf{\bar{l}_{1}}$ and $\mathbf{\bar{l}_{2}}$ in
$\chi$PT: choice of the most stringent point}
\label{sec:ChiPT}

It is simple to convert the conditions displayed in Eq.~(\ref{eq:positive}) into bounds for chiral LECs. The region $\mathcal{A}$ covers a very low energy domain, and is below the $2\,\pi$ threshold in any of the three crossed channels. In this range of energies one expects the chiral expansion to work well, so we will approximate the right-hand side of Eq.~(\ref{eq:positive}) by the $\chi$PT result at $\mathcal{O}(p^{4})$.

Since the $\chi$PT amplitude is derived from a local Lagrangian, it automatically respects  the principles of crossing symmetry, unitarity, and analyticity. One could na\"\i vely argue that the positivity constraints should also be automatically satisfied, but this is not necessarily true. As noted in Ref.~\cite{Ananthanarayan:1994hf},  $\chi$PT is an expansion in low momenta, so the amplitude has polynomial behavior (up to logarithms) and grows as $s^2$ or even worse at higher orders, violating the Froissart bound. As a result, the positivity constraints provide additional information beyond $\chi$PT, and give restrictions on the LECs.

The $\chi$PT leading order amplitude is linear in $s$ and $t$ and so vanishes on taking the second derivative; the next-to-leading order amplitude does not. The $\mathcal{O}(p^{4})$ amplitude can be found in Ref.~\cite{Gasser1}, and its second derivative depends only on two LECs: $\bar{l}_{1}$ and $\bar{l}_{2}$ in the $SU(2)_L \times SU(2)_R$ chiral Lagrangian. The amplitude can be split into polynomial terms quadratic in momenta and masses, and chiral logarithms. The former contain the LECs and their second derivatives yield energy independent terms; the latter
depend only on momenta and masses, are independent of the order $p^4$ LECs, and give energy dependent contributions to the second derivative. The general structure of the bound can thus be written as 
\begin{equation}
\sum_{i=1}^{2}\alpha_{ji}\,\bar{l}_{i}\,-\, f_{j}[\,(s,t)\in\mathcal{A}\,]\,\ge\,0\quad j\,=\,1,2,3\,,\label{eq:chiral_bound}
\end{equation}
where $\alpha_{ji}$ are real coefficients and $f_{j}(s,t)$ are functions 
obtained from chiral logarithms and LEC-independent polynomial
terms, and $j$ labels each one of the processes in Eq.~(\ref{eq:positive}). The most stringent restriction is obtained for those values of $(s,t)$
that maximize $f_{j}(s,t)$ inside the region $\mathcal{A}$:
\begin{equation}
\sum_{i=1}^{2}\alpha_{ji}\,\bar{l}_{i}\,\ge\,f_{j}[\,(s,t)\in\mathcal{A}\,]
\,\Bigr|_{\mathrm{max}}\,.\label{eq:max_bound}
\end{equation}

It is important to estimate possible corrections to the bounds in Eq.~(\ref{eq:max_bound}) coming from $\mathcal{O}(p^6)$ terms in the amplitude. The computation of the $\pi\,\pi$ scattering amplitude at this level of precision was performed in Ref.~\cite{Bijnens:1995yn}, and can be split into three pieces: two-loop terms (double chiral logarithms), that only depend on $m$ and $F_\pi$; one-loop terms (single chiral logarithms), that depend linearly on several $\mathcal{O}(p^4)$ LECs (not only $l_1$ and $l_2$); and tree-level terms, that depend on $\mathcal{O}(p^6)$ LECs. In Ref.~\cite{Dita:1998mh}, Eq.~(\ref{eq:chiral_bound}) was calculated  with the corresponding $\mathcal{O}(p^6)$ amplitude for $\pi^{0}\pi^{0}$ and $\pi^{+}\pi^{0}$ at $s=0$, $t=4\,m^2$. Unfortunately the corresponding $\mathcal{O}(p^6)$ LECs are badly known (resonance saturation estimates are usually used), and the chiral LECs we want to bound, $l_1$ and $l_2$, appear again in the one-loop terms. In addition the rest of LECs in the one-loop terms are symmetry breaking operators, and hence appear always multiplied by the pion mass. As a result, their numerical values are poorly known. So we have only control over the two-loop terms. To get an educated guess for the error from the $\mathcal{O}(p^6)$ terms, we will multiply the value of the purely two-loop correction by a factor of 3. To be more conservative we will adopt as a common error for the three bounds, the biggest of these, which is $0.4$.

There is one last issue to be discussed before we show our results. It is well known that the scalar one-loop two-point function is not smooth at threshold (for instance its imaginary part is zero below threshold but nonzero above). Its first and second derivatives tend to infinity when we approach threshold from below. So in order for the positivity condition to hold, the coefficients in front of these first and second derivatives must always be positive below threshold. This is indeed the case in all processes under study in our work.

\begin{figure}
\begin{center}
\includegraphics[width=8cm]{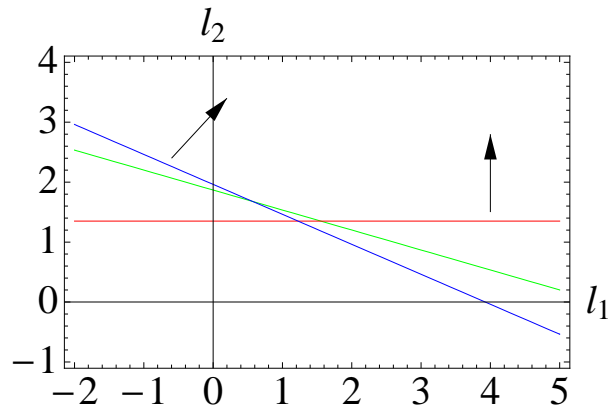}
\end{center}
\caption{The $\bar{l}_{1}-\bar{l}_{2}$ region allowed by the positivity conditions is shown. The three lines correspond to the three bounds in Table~\ref{tab:bounds}. We also show the experimentally fitted values of Ref.~\cite{Colangelo} with their error.\label{fig:graphic-bound}}
\end{figure}

We find that the maximum of $f_j(s,t)$ is always achieved for $t=4\, m^{2}$, regardless of the process (i.e.\ for $j=1,2,3$); the value for $s$ does depend on the particular process.  The maximum of $f_j$ is at $s=0$ for $j=1,2$. For $j=3$, the maximum was found numerically to be at $s=1.114 \,m^2$. Our results are summarized in  Table~\ref{tab:bounds} together with  a comparison with the values for the experimentally fitted LECs $\bar{l}_{1}=-0.4\pm0.6$ and $\bar{l}_{2}=4.3\pm0.1$ from Ref.~\cite{Colangelo}.  In Fig.~\ref{fig:graphic-bound}, we plot the allowed region in the $\bar{l}_{1}-\bar{l}_{2}$ space parameter, together with the experimentally fitted value.
\renewcommand{\arraystretch}{1.5}
\begin{table*}
\begin{center}\begin{tabular}{|c|c|c|c|c|}
\hline 
Process& LECs  & Maximum position & Bound & Fit to expt.\tabularnewline
\hline
$\pi^{0}\pi^{0}\to\pi^{0}\pi^{0}$&
$\bar{l}_{1}+2\,\bar{l}_{2}$&
$(s=0,\, t=4\,m^{2})$&
$\ge\frac{157}{40}=3.925 \pm 0.4 $&
$\phantom{5}8.2\pm0.6$\tabularnewline
$\pi^{+}\pi^{0}\to\pi^{+}\pi^{0}$&
$\bar{l}_{2}\phantom{+2\,\bar{l}_{2}}$&
$(s=0,\, t=4\,m^{2})$&
$\,\,\ge\frac{27}{20}=1.350 \pm 0.4 $&
$\phantom{5}4.3\pm0.1$\tabularnewline
$\pi^{+}\pi^{+}\to\pi^{+}\pi^{+}$&
$\bar{l}_{1}+3\,\bar{l}_{2}$&
$(s=1.114\, m^{2},\, t=4\,m^{2})$&
$\phantom{\frac{157}{40}}\,\,\,\,\,\ge5.604 \pm 0.4$&
$12.5\pm0.7$\tabularnewline
\hline
\end{tabular}\end{center}
\caption{Bounds obtained by unitarity, crossing, and analyticity and comparison with values extracted from
a fit to the experimental data given in Ref.~\cite{Colangelo}. The error on the bound is an estimate of the order $p^6$ terms.\label{tab:bounds}}
\end{table*}

\section{Comparison with previous analyses}
\label{sec:Comparison}

As mentioned in Sec.~\ref{sec:Introduction}, there are several studies in the literature that combine $\chi$PT with axiomatic principles. In this section we compare with previous results and point out the advantages of the method used here. 

In Ref.~\cite{Ananthanarayan:1994hf} only the $\pi^0\pi^0$ amplitude was considered, and so only bounds on $\bar{l}_{1}+2\,\bar{l}_{2}$ could be obtained. From the requirement that the $s$-wave amplitude has a minimum in the interval $1.217\le s/m^2 \le 1.697$ they obtain $\bar{l}_{1}+2\,\bar{l}_{2}\ge 3.32 \pm 0.85 $. This value is less restrictive than our bound,  and in addition has a much bigger uncertainty. From the once subtracted dispersion relation of the full  $\pi^0\pi^0$ amplitude they obtain $\bar{l}_{1}+2\,\bar{l}_{2}\ge 3.3 \pm 2.5 $, which has a very large error and is weaker than our bound.  Using the Froissart-Gribov representation for the $d$-wave partial amplitude, they obtained our value for the bound, but since a reliable estimate of its error was not found, this result was not taken into account in the final results in Ref.~\cite{Ananthanarayan:1994hf}.

In Ref.~\cite{Pennington:1994kc}, the Froissart-Gribov representation for the $d$-wave scattering lengths was used to derive positivity conditions. In this way, they obtained the same results as us for $\bar{l}_{1}+2\,\bar{l}_{2}$ and $\bar{l}_{2}$, with no errors quoted. In Appendix~A we demonstrate that this method is equivalent to ours for the particular point $s=0$, $t=4\,m^2$.

In Ref.~\cite{Dita:1998mh}, the analysis of Ref.~\cite{Ananthanarayan:1994hf} was repeated, requiring a minimum of the $s$-wave amplitude in the same interval as above, $1.217\le s/m^2 \le 1.697$. Surprisingly Ref.~\cite{Dita:1998mh} obtained a much more stringent bound, $\bar{l}_{1}+2\,\bar{l}_{2}\ge 6.16$ (no error quoted). In view of the discussion in both papers, it is our belief that Ref.~\cite{Ananthanarayan:1994hf} gives the correct answer. The main analysis of Ref.~\cite{Dita:1998mh} uses the same method that we do, and in the same domain $\mathcal{A}$. It is argued that the most  stringent point necessarily lies on the $2\,s+t=4\,m^2$ line, but we do not see why this should be true. In fact, we explicitly find that for the $\pi^+\pi^+$ amplitude, it is not on this line. Furthermore, Ref.~\cite{Dita:1998mh} only displays the bounds at $s=0$ ($t=4\,m^2$), where we get the same results for $\bar{l}_{1}+2\,\bar{l}_{2}$ and $\bar{l}_{2}$, and at $s=-\,4\,m^2$ ($t=12\,m^2$), where the bounds are much more restrictive.  The result $\bar{l}_{1}+\bar{l}_{2}\ge 4.914$ quoted in Ref.~\cite{Dita:1998mh}
at $s=-\,4\,m^2$ ($t=12\,m^2$) is violated by the experimentally fitted values of Ref.~\cite{Colangelo}. Even though Ref.~\cite{Dita:1998mh} uses the same domain $\mathcal{A}$ as our analysis, for the numerics, they trespass outside this region. The bounds $\bar{l}_{1}+2\,\bar{l}_{2}\ge 6.923$, $\bar{l}_{2}\ge 2.01$, and $\bar{l}_{1}+\bar{l}_{2}\ge 4.914$ obtained in Ref.~\cite{Dita:1998mh} at $s=-\,4\,m^2$ ($t=12m^2$) should not be trusted since the fixed-$t$ dispersion relations are not valid for $t>4\,m^2$.

Finally, in Ref.~\cite{Distler:2006if} the same method of Ref.~\cite{Dita:1998mh} is used, but only in the Mandelstam triangle, which is why their bound for $\bar{l}_{1}+3\,\bar{l}_{2}$ is less restrictive than ours, and does not exclude any values for $\bar{l}_{1,2}$ not already excluded by the bounds on $\bar{l}_{1}+\,\bar{l}_{2}$
and $\bar{l}_{2}$.

\section{Unitarity relations for the linear sigma model}
\label{sec:sigma-model}

In Sec.~\ref{sec:ChiPT}, we substituted the $\chi$PT results into Eq.~(\ref{eq:positive}) and obtained bounds on some undetermined low-energy constants in the effective Lagrangian. One can repeat this exercise for theories in which the low-energy effective Lagrangian is calculable, to test the validity of the bounds. In this section we perform such an analysis for the linear sigma model. 

The most straightforward method is to use the predictions of the LSM for $\bar{l}_{1}$ and $\bar{l}_{2}$ for the bounds displayed in Table~\ref{tab:bounds}. As already explained in Sec.~\ref{sec:Introduction}, the LSM is invariant under the same symmetries as $\chi$PT, and so all operators obtained after integrating out the $\sigma$ particle must belong to the $\chi$PT Lagrangian at some order in the chiral expansion. In Ref.~\cite{Gasser1} this computation was performed at the one-loop level and at $\mathcal{O}(p^{4})$, the following result was obtained:
\begin{eqnarray}
\bar{l}_{1}&=&\frac{24\,\pi^{2}}{g}+2\,\log\left(\frac{m_{\sigma}}{m}\right)-
\frac{35}{6},\nn
\bar{l}_{2}&=&2\,\log\left(\frac{m_{\sigma}}{m}\right)-\frac{11}{6}\,,
\label{eq:LECLSM}
\end{eqnarray}
leading to the inequalities
\begin{eqnarray}
\frac{24\,\pi^{2}}{g}\,+\,6\,\log\left(\frac{m_{\sigma}}{m}\right)&\ge&\frac{537}{40}\,,\nn
\log\left(\frac{m_{\sigma}}{m}\right)&\ge&\frac{191}{120}\,,\nn
\frac{24\,\pi^{2}}{g}\,+\,8\,\log\left(\frac{m_{\sigma}}{m}\right)&\ge&16.94\,.
\label{eq:drama}
\end{eqnarray}
where $g$ is the (weak) coupling constant of the $\phi^4$ term in the LSM. It can be written (at leading order) in terms of the pion decay constant $F_{\pi}$ through the relation $2\,g= (m_{\sigma}^{2}-m^2)/F_{\pi}^{2}$.\footnote{Note that in Ref.~\cite{Gasser1} the relation $2\,g= m_{\sigma}^{2}/F_{\pi}^{2}$ is used.  Instead, we identify $F_\pi$ with the vacuum expectation value of the $\sigma$ field $v$, which coincides with the pion decay constant at leading order. In the nonlinear parametrization of the LSM,  the pion fields are collected in the exponential matrix $\exp(i\, \bm{\tau}\cdot\bm{\pi}/v)$, which coincides with the $\chi$PT form after the identification $F_\pi=v$ is made. In addition, $m_{\sigma}$ depends on the pion mass but at leading order the combination $m_{\sigma}^{2}-m^2$ does not. Although in practice the choice of $F_\pi$ does not affect the results of Eq.~(\ref{eq:drama}) and the discussion in this section, we prefer to use the notation of Ref.~\cite{Bessis:1972sn}.} These results are obtained in weak-coupling perturbation theory to one loop, and have corrections of order $g$ from the two-loop graphs. The first and third relations of Eq.~(\ref{eq:drama}) are always satisfied for a weakly coupled theory on which Eq.~(\ref{eq:LECLSM}) relies, since the $24\,\pi^2/g$ term is larger than the other terms for small values $g$. Note that the coefficient of the $1/g$ term must have the correct sign for the inequality to be satisfied, which it does. The second relation does not involve an inverse power of the coupling constant, and is not satisfied for large enough values of $m/m_\sigma$. In particular, it is violated if $m_{\sigma}\lesssim4.9\, m$. One way out of this contradiction is that the derivation of the inequality, which relies on the Froissart bound, is not valid. But it is not difficult to show that the LSM is a local renormalizable theory and satisfies the Froissart bound.  In the chiral limit $m \to0$ and the bound is satisfied. The Goldstone boson is made massive by a symmetry breaking term (analogous to an external magnetic field). The strength of the symmetric breaking term must be increased to increase $m$. The symmetry breaking term also contributes to the $\sigma$ mass, so another way out is if the region $m_\sigma/m \lesssim 4.9$ is not possible for any values of the parameters in the LSM. By explicitly computing the masses in the LSM with a symmetry breaking term, one can show that any $m_\sigma/m \ge \sqrt{3}$ is allowed, and since $\sqrt 3 < 4.9$ there are allowed values for the mass ratio which violate the bound.

The loophole in the argument is that for low values of the $\sigma$ mass, the higher $1/m_{\sigma}^{2}$ corrections become more important. Results in Table~\ref{tab:bounds} rely on the fact that in $\chi$PT, the scattering amplitude can be safely truncated at $\mathcal{O}(p^{4})$, which translates into the statement that the LSM amplitude can be truncated at $\mathcal{O}(m_{\sigma}^{-2})$. If $m_{\sigma}$ is not big enough, this approximation receives sizable corrections and the chiral expansion breaks down. To violate the bound in the second of Eqs.~(\ref{eq:drama}) requires $m_\sigma \lesssim 4.9. 1\, m$. The chiral expansion is formally an expansion in powers of $m/m_\sigma$, and the bound is violated when $m^2/m_\sigma^2 \agt 0.04$, a finite distance away from the origin. What is surprising is that this number, which is formally of order unity, is numerically much smaller than one would have naively guessed.

\begin{figure}
\begin{center}\includegraphics[width=7cm]{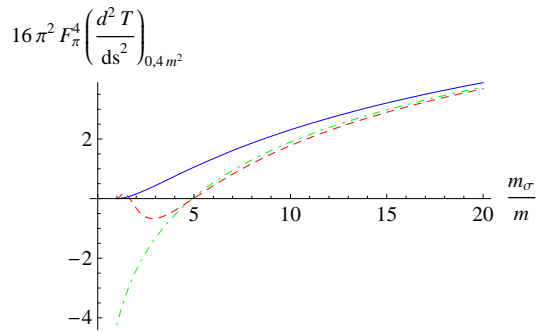}
\end{center}
\caption{Plot of $16\,\pi^{2} F_\pi^{4}\,\mathrm{d}^{2}T(s,4\,m^{2})/\mathrm{d}s^{2}\bigr|_{s=0}$ in the linear sigma model for the $\pi^{+}\pi^{0}\to\pi^{+}\pi^{0}$ process as a function of $m_{\sigma}/m$. The exact amplitude (blue, continuous line) is positive for $m_{\sigma}>m$. The amplitude up to and including $1/m_\sigma^4$ terms (red, dashed line), is positive for $m_{\sigma}>5\,m$. The $\mathcal{O}(m_\sigma^{-2})$ amplitude (green, dot-dashed line) 
remains negative for $m_\sigma<4.9\,m$.
\label{fig:sigma-positive}}
\end{figure}

As a first approach, we include the $1/m_{\sigma}^{4}$ corrections to the amplitude and find that then the bounds
are violated for $m_{\sigma}\lesssim5\, m$, which is not 
satisfactory, but indicates that the
$1/m_{\sigma}^{2}$ expansion is slowly converging, and the $1/m_\sigma^4$ term contribution moves the result in
the direction of restoring the validity of the bound, since it makes the second derivative of the amplitude less negative (see
Fig.~\ref{fig:sigma-positive}). To test the LSM bound we will apply directly
Eq.~(\ref{eq:positive}), rather than  the expanded form Eq.~(\ref{eq:LECLSM}), to the LSM scattering amplitude
prediction for the $\pi^{+}\pi^{0}\to\pi^{+}\pi^{0}$ process. The second derivative of the tree-level amplitude for
this process within the LSM vanishes, and so one needs the one-loop result. In Ref.~\cite{Bessis:1972sn} this
calculation was performed using a mass-dependent subtraction scheme. The result is expressed in terms of finite
two-, three-, and four-point scalar one-loop integrals, which are then expanded in inverse powers of
$m_{\sigma}^{2}$. We will use instead the numerical values for the full integral expressions. The renormalization
procedure followed in Ref.~\cite{Bessis:1972sn} is perfectly acceptable for our computation, since the physical
amplitudes are scheme independent. Most modern computations are done in the 
$\mathrm{\overline{MS}}$ scheme. In Appendix~B we give the one-loop LSM amplitudes in the mass-independent $\mathrm{\overline{MS}}$ scheme, a result which does not appear in the literature.

Recently, in Ref.~\cite{Bissegger:2007ib}, the leading logarithms of the scalar-scalar QCD Green function have been
calculated to two-loop accuracy. For the renormalization the (modified) $\mathrm{\overline{MS}}$ scheme is used, 
as we do in Appendix~B. However they assume the chiral limit (massless pions) and small external momenta, which is 
equivalent to expanding in inverse powers of $m_\sigma$, approach already followed in Ref.~\cite{Bessis:1972sn}. For our analysis we need a nonzero pion mass and arbitrary values of the external momenta. Nevertheless, 
the renormalization program is mass independent and so in both calculation should coincide. We agree with their 
results.

The second derivative ${\mathrm{d}^{2}}T(s,4\, m^{2})/{\mathrm{d}s^{2}}|_{s=0}$ for the $\pi^{+}\pi^{0}\to\pi^{+}\pi^{0}$ process in the LSM is computed for any value of the $m_{\sigma}/m$ ratio. The results are shown in Fig.~\ref{fig:sigma-positive}, which clearly shows that the positivity condition is satisfied at the one-loop level in the LSM for any value of the $\sigma$ mass bigger than the pion mass (even though it would suffice to be satisfied for $m_{\sigma}>\sqrt{3}\, m$). The apparent contradiction of Eq.~(\ref{eq:drama}) was only due to the poor convergence of the $1/m_{\sigma}^{2}$ expansion of the LSM amplitude for small $m_{\sigma}$. The nonlinear sigma model (understood as the nonrenormalizable effective field theory obtained by integrating the $\sigma$ field out the LSM action) 
is consistent (i.e.\ obeys the axiomatic bounds) only if we 
(at least) include  the $\mathcal{O}(p^8)$ contribution.

This should serve as a warning for the estimate of chiral LECs by resonance saturation.  In such determinations, one starts with a  chiral invariant Lagrangian with resonances as explicit degrees of freedom~\cite{RChT}. The values of the chiral LECs are obtained, as in the LSM, by functionally integrating out the hadronic resonances. The ratio $m_\rho/m\sim5.5$ is of the same order than the value $m_\sigma/m$ that makes the LSM chiral expansion fail. However, we believe that since in the Lagrangians of \cite{RChT}, all  LECs are already generated at tree level, this anomalous behavior is absent. In the LSM, $l_2$ is only generated at one-loop, which is why the middle inequality in Eq.~(\ref{eq:drama}) does not have a $1/g$ term and has poor convergence in the $1/m_\sigma^2$ expansion.

At this point a natural question arises. Since $m_K/m\sim3.5< 5$ it could be inconsistent to integrate the kaon out of the $SU(3)$ $\chi$PT action to obtain the $SU(2)$ chiral LECs. In fact using the $L_1,\,L_2$,  and $L_3$ values of Ref.~\cite{Amoros:2001cp}, we obtain $\bar{l}_1=5.64\pm0.84$ and $\bar{l}_2=1.95\pm0.23$ which do not agree well with the values quoted in Ref.~\cite{Colangelo}, but are in agreement with our bounds. The additional complications that arise on imposing the positivity conditions to $SU(3)$ $\chi$PT are discussed in another paper~\cite{Mateu:2008gv}.

\section{Conclusions}
\label{sec:Conclusions}

There are nontrivial constraints which follow from unitarity, analyticity, and crossing symmetry which must be satisfied by any relativistic quantum theory. There are some interesting and nontrivial constraints on low-energy effective theories which arise by imposing these constraints on the effective theory scattering amplitude.

In this work we have transformed the dispersion relations for the $\pi\,\pi$ scattering amplitude into positivity conditions for several processes, valid in a certain region of the Mandelstam plane below threshold. This region is in fact larger than the Mandelstam triangle, as commonly assumed. These positivity conditions can be converted into bounds for two LECs of the $SU(2)$ $\chi$PT Lagrangian.  Our analysis leads to a stronger bound than those obtained previously, since we use positivity in a larger region of the Mandelstam plane. The values of the LECs extracted from experiment are consistent with the bounds derived in this paper.

One  nice feature of the structure of the bounds is that it correlates two distinct pieces of the $\mathcal{O}(p^4)$ amplitude: LECs and chiral logarithms. Whereas the former is leading order in the $1/N_C$ counting and represents an expansion in $1/m_\rho^2\sim (0.7\,\mathrm{GeV})^{-2}$, the  latter is subleading in large-$N_C$ and represents an expansion in  $1/\Lambda_\chi^2\sim (1.1\, \mathrm{GeV})^{-2}$ where $\Lambda_\chi\sim 4\,\pi F_\pi$ and $F_\pi$ is the decay constant of the pion~\cite{nda}.

One can use Eq.~(\ref{eq:disp_final}) with $n=4$ to obtain bounds for higher order LECs, using the amplitude up to order $\mathcal{O}(p^6)$. The $\mathcal{O}(p^4)$ LECs in the $\mathcal{O}(p^4)$ amplitude vanish on taking the fourth derivative but the one-loop $\mathcal{O}(p^4)$ chiral logarithmic terms do not. However, one-loop diagrams with one insertion of the $\mathcal{O}(p^4)$ LECs contribute to terms of order $p^6$ times chiral logarithms, which do not vanish on taking the fourth derivative. The $\mathcal{O}(p^6)$ LECs also contribute to the fourth derivative. Thus one now gets  inequalities involving the $\mathcal{O}(p^4)$ LECs and $\mathcal{O}(p^6)$ LECs plus $\mathcal{O}(p^4)$ chiral  logarithms. In addition to having a lot of LECs we no longer compare terms of the same order in the chiral expansion.

The low-energy limit of the linear sigma model Lagrangian is a theory with spontaneous chiral symmetry breaking in which the LECs can be computed in terms of the coupling constant $g$ of the LSM.  The values of $\bar{l}_{1}$ and $\bar{l}_{2}$ for this model are in apparent violation of the positivity bounds for $m_{\sigma}\lesssim4.9\, m$, while the range $m_{\sigma}>\sqrt{3}\, m$ can be realized in the LSM. We have shown that the apparent violation is an artifact of the truncation of the $1/m_{\sigma}^{2}$ corrections and that the LSM is consistent with the positivity conditions
for $m_{\sigma}\ge m$.

In a subsequent work~\cite{Mateu:2008gv} we will apply the same method to the $SU(3)$ case, including the full octet of pseudo-Goldstones and generalize the method to take into account the explicit breaking of $SU(3)_{V}$ symmetry. This gives bounds on $L_{1}$, $L_{2}$, and $L_{3}$.

\section*{Acknowledgments}

We are grateful to professor B.~Ananthanarayan for making us aware of the literature on this field and for various and useful comments. We would also like to thank G.~Wanders for some helpful correspondence and J.~J.~Sanz-Cillero for pointing out several typos in earlier versions of the manuscript. V.~M. thanks the University of California at San Diego for its hospitality, where the major part of this work was done. The work of V.~Mateu is supported by a FPU contract (MEC). This work has been supported in part by the EU MRTN-CT-2006-035482 (FLAVIAnet), by MEC (Spain) under Grant No. FPA2004-00996 and by Generalitat Valenciana under Grant No. GVACOMP2007-156.

\begin{appendix}

\section{Relation between our method and scattering lengths}

In this appendix we wish to demonstrate how the procedure followed in Ref.~\cite{Pennington:1994kc} is related to ours. Let us start by recalling the definition of the scattering lengths. From the partial-wave decomposition of Eq.~(\ref{eq:partial_s}), one defines for each spin and isospin amplitude the scattering lengths $a_{\ell}^{I}$
\begin{equation}
a_{\ell}^{I}\,=\,\lim_{s\to4\,m^{2}}\frac{f_{\ell}^{I}(s)}{\left(\frac{s}{4}-m^{2}\right)^{\ell}}\,.
\end{equation}
For even $\ell$, the $I=1$ scattering length must vanish because of Bose symmetry. In Ref.~\cite{Pennington:1994kc} these scattering lengths can be shown to satisfy the positivity conditions
\begin{equation}
a_{2}^{0}\,+\,2\, a_{2}^{2}\,\ge\,0\,,\qquad a_{2}^{0}\,-\, a_{2}^{2}\,\ge\,0\,.
\label{eq:Portoles}
\end{equation}
using the Froissart-Gribov representation.

\noindent It is not difficult to relate the scattering lengths to the $\ell$-derivative of the total spin-$I$ scattering amplitude:
\begin{eqnarray}
a_{\ell}^{I}&=&\dfrac{4^{\ell}\,\ell\,!}{(2\,\ell+1)}\left.
\frac{\mathrm{d}^{\ell}T^{I}(4\, m^{2},t)}{\mathrm{d}\, t^{\ell}}
\right|_{t=0}\nn
&=&\dfrac{4^{\ell}\,\ell\,!}{(2\,\ell+1)}\, C_{t}^{II^{\prime}}
\left.\frac{\mathrm{d}^{\ell}T^{I^{\prime}}(s,4\, m^{2})}{\mathrm{d}\, s^{\ell}}
\right|_{s=0}\,,\label{eq:lenght-der}
\end{eqnarray}
where we have used a relation analogous to Eq.~(\ref{eq:crossing-u})
\begin{eqnarray}
T^{\, I}(s,t)&=& C_{t}^{II^{\prime}}T^{\, I^{\prime}}(t,s)\,,\nn
C_{t}^{II^{\prime}}C_{t}^{I^{\prime}J}&=&\delta_{IJ}\,,
\nn
C_{t}&=&\frac{1}{6}\left(\begin{array}{ccc}
2 & \phantom{+\,}6 & \phantom{+}10\\
2 & \phantom{+}3 & -\,5\\
2 & -\,3 & \phantom{+\,}1\end{array}\right)\,,\label{eq:crossing-t}
\end{eqnarray}
which follows from crossing symmetry in the $t$-channel.

For even $\ell$ and $I=1$, Eq.~(\ref{eq:lenght-der}) implies that the corresponding scattering length is identically zero. To see this, recall that $T^{1}(4\, m^{2},t)=-\, T^{1}(4\, m^{2},-\,t)$ by Bose symmetry. Now, since the point $s=0$, $t=4\, m^{2}$ lies in the region $\mathcal{A}$, for $\ell=2$ we know that certain linear combinations of the derivatives appearing in the last equality of Eq.~(\ref{eq:lenght-der}) must be positive. Inverting Eq.~(\ref{eq:lenght-der}) we obtain
\begin{equation}
\left.\frac{\mathrm{d}^{2}\,T^{I}(4\, m^{2},t)}{\mathrm{d}\, t^{2}}
\right|_{t=0}\,=\,\frac{5}{32}\, C^{IJ}_t a^I_2\,.
\end{equation}
Using the linear combinations that give the amplitudes in  Eqs.~(\ref{eq:positive}), and bearing in mind that $a_{2}^{1}\equiv0$, we immediately reproduce the result shown in Eq.~(\ref{eq:Portoles}) plus the linearly dependent relation $2\,a_{2}^{0}+a_{2}^{2}\ge0$.

We have demonstrated that the method in Ref.~\cite{Pennington:1994kc} corresponds to using positivity at the $s=0$, $t=4\,m^2$ point in region $\mathcal{A}$ of the Mandelstam plane. This is why Ref.~\cite{Pennington:1994kc} did not find our third bound, which arises from $s=1.114\,m^2$, $t=4\,m^2$.

\section{One-loop $\pi\,\pi$ scattering in the linear
sigma model}

In this appendix,  we renormalize the linear sigma model at one loop for finite pion mass. We will use the mass-independent $\mathrm{\overline{MS}}$ scheme, instead of the subtraction scheme of Ref.~\cite{Bessis:1972sn}. So our Lagrangian will be split  into renormalized pieces and counterterms. The basic building block containing all fields in the LSM is the $SU(2)$ matrix 
\begin{equation}
\widetilde{\Sigma}\,=\,\tilde{\sigma}\,+\, i\,\bm{\pi}\cdot\bm{\tau}\,=\,
v\,+\,\sigma\,+\, i\,\bm{\pi}\cdot\bm{\tau}\,\equiv\, v\,+\,\Sigma\,,
\end{equation}
where $\bm{\tau}$ are the three Pauli matrices, $\sigma=\tilde{\sigma}-v$ is the sigma field with zero vacuum expectation value (VEV), and $v=\langle\tilde{\sigma}\rangle$. The Lagrangian is
\begin{eqnarray}
\mathcal{L}^{\mathrm{LSM}}&=&\frac{1}{4}\left\langle \partial_{\mu}\widetilde{\Sigma}\,
\partial^{\mu}\widetilde{\Sigma}^{\dagger}
\right\rangle +\frac{\mu^{2}}{4}\left\langle \widetilde{\Sigma}\,\widetilde{\Sigma}^{\dagger}
\right\rangle -\frac{g}{16}\left\langle \widetilde{\Sigma}\,\widetilde{\Sigma}^{\dagger}
\right\rangle ^{2}\nn
&&+\,\beta\left\langle \widetilde{\Sigma}+\widetilde{\Sigma}^{\dagger}\right\rangle  + \mathcal{L}_{c.t.}\,.
\end{eqnarray}
At leading order we get the following relations for masses and VEV:
\begin{eqnarray}
m^{2}&=&\frac{4\,\beta}{v}\,,\nn
m_{\sigma}^{2}&=&2\, g\, v^{2}+\frac{4\,\beta}{v}\,=
\,2\, g\, v^{2}+m^{2}\,,\nn
v^{2}&=&\frac{m_{\sigma}^{2}-m^{2}}{2\, g}\,.
\end{eqnarray}
\begin{figure}[t]
\begin{center}
\includegraphics[width=6cm]{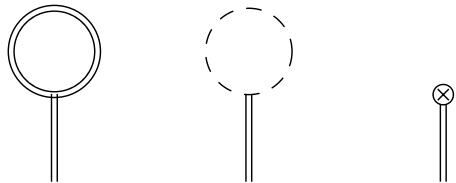}
\end{center}
\caption{One-point $\sigma$ function. Double and dashed lines denote $\sigma$ particles and pions, respectively.
The sum of all these tadpole graphs must vanish to ensure that perturbation theory is done around the minimum of the potential, including quantum corrections.
\label{fig:One-point}}
\end{figure}
At tree level, one can calculate the VEV directly from the Lagrangian by minimizing the potential. At one-loop, the most convenient procedure is to impose the condition that the one-point $\sigma$ function identically vanishes, as shown in Fig.~\ref{fig:One-point}. 
This ensures that we are considering quantum excitations around an extremum of the potential. It also implies that one-point functions (tadpoles) are zero in any graph, so we will not display this topology.
\begin{figure*}
\begin{center}
\includegraphics[width=12cm]{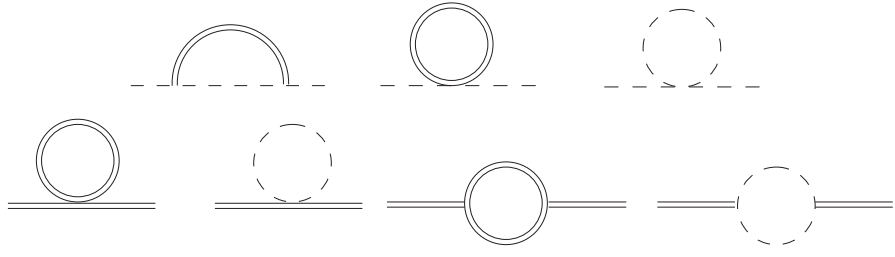}
\end{center}
\caption{One-loop corrections to the pion (top) and $\sigma$ (bottom) propagators.\label{fig:propagators}}
\end{figure*}

Let us start calculating the quantum corrections for the $\pi$ and $\sigma$ propagators, as shown in Fig.~\ref{fig:propagators}. The pion propagator is diagonal in isospin, and thus proportional to $\delta^{ab}$, which we drop. The renormalized one-loop contributions are
\begin{eqnarray}
T^{\pi} & = & \frac{g}{16\,\pi^{2}}\left[2\left(m_{\sigma}^{2}-m^{2}\right)I_{\sigma\pi}(q^{2})-
2\, m_{\sigma}^{2}\, A_{\sigma}+2\, m^{2}A_{\pi}\right]\,,\nonumber \\
T^{\sigma} & = & \frac{3\,g}{16\,\pi^{2}}\left(m_{\sigma}^{2}-m^{2}\right)\left[I_{\pi\,\pi}(q^{2})+
3\,I_{\sigma\sigma}(q^{2})\right]\,,
\end{eqnarray}
where we have defined
\begin{eqnarray}
A_{\pi}&=&1-\log\left(\frac{m^{2}}{\mu^{2}}\right),\quad A_{\sigma}\,=\,1-\log\left(\frac{m_{\sigma}^{2}}{\mu^{2}}\right),\nn
I_{\pi\,\pi}(q^{2}) & = & A_\pi+1-\beta_{\pi}(q^{2})\log
\left(\frac{\beta_{\pi}(q^{2})+1}{\beta_{\pi}(q^{2})-1}\right)\,,\nonumber \\
I_{\sigma\sigma}(q^{2}) & = & A_\sigma+1-\beta_{\sigma}(q^{2})
\log\left(\frac{\beta_{\sigma}(q^{2})+1}{\beta_{\sigma}(q^{2})-1}\right)\,,\nonumber \\
I_{\sigma\pi}(q^{2})&=&1+\dfrac{1}{m_\sigma^{2}-m^{2}}\left[m_{\sigma}^{2}\,A_\sigma-m^{2}\,A_\pi\right]\nn
&&+\,\dfrac{1}{2}\left(\dfrac{m_{\sigma}^{2}-m^{2}}{s}-\dfrac{m_{\sigma}^{2}+m^{2}}{m_{\sigma}^{2}-m^{2}}\right)\log\left(\dfrac{m^{2}}{m_{\sigma}^{2}}\right)\nn
&&-\,\dfrac{\nu(s)}{2\, s}\log\left\{ \dfrac{[s+\nu(s)]^{2}-(m_{\sigma}^{2}-m^{2})^{2}}{[s-\nu(s)]^{2}-(m_{\sigma}^{2}-m^{2})^{2}}\right\},\nn
\beta_{\pi}(q^{2}) & = & \sqrt{1-\frac{4\,m^{2}}{q^{2}}}\,,\qquad\beta_{\sigma}(q^{2})\,=
\,\sqrt{1-\frac{4\,m_{\sigma}^{2}}{q^{2}}}\,,\nn
\nu(s)&=&\sqrt{[s-(m_{\sigma}^{2}+m^{2})][s-(m_{\sigma}^{2}-m^{2})]},
\end{eqnarray}
From the renormalization of the propagators one can obtain the running of the $g$ coupling constant
\begin{equation}
\dfrac{\mu}{g}\,\dfrac{\mathrm{d}g}{\mathrm{d}\mu}\,=\,\dfrac{3}{2\,\pi^{2}}\, g\,,\label{eq:running}
\end{equation}
which ensures that observables are $\mu$-independent.

\begin{figure*}
\begin{center}
\includegraphics[width=12cm]{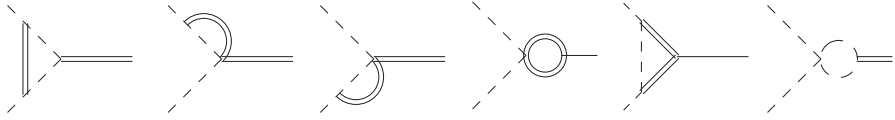}
\end{center}
\caption{Vertex quantum corrections to the $\sigma-\pi\,\pi$ interaction.\label{fig:Vertex}}
\end{figure*}

Next we calculate the vertex correction to the $\sigma\,\pi\,\pi$ interaction, that is, the irreducible three-point function. This correction would affect, among other things, the decay of the $\sigma$ into two pions. The diagrams are shown in Fig.~\ref{fig:Vertex}. Since the $\sigma$ is an isospin singlet, its coupling to the pair $\pi^{a}\,\pi^{b}$ must be proportional to $\delta^{ab}$ which again will not be displayed. The renormalized result then reads
\begin{eqnarray}
T^{\sigma-\pi\,\pi} & = & -\,2\, g\, v+\frac{g^{2}v}{8\,\pi^{2}}\Biggl[2\left(m_{\sigma}^{2}-m^{2}\right)V_{\sigma}(s)\nn
&&+\,
6\left(m_{\sigma}^{2}-m^{2}\right)V_{\pi}(s)
+  4\, I_{\sigma\pi}(m^{2})\nn
&&+\,5\, I_{\pi\,\pi}(s)+3\, I_{\sigma\sigma}(s)\Biggr]\,,
\end{eqnarray}
where we define three-point one-loop functions as
\begin{widetext}
\begin{eqnarray}
V_{\sigma}(s) & = & -\,\frac{1}{s\,\beta_{\pi}(s)}\Biggl\{ 2\,\mathrm{Li}_{2}\left[\frac{4\, m^{2}-2\, m_{\sigma}^{2}+s\left(\beta_{\pi}(s)-1\right)}{8\, m^{2}-2\, m_{\sigma}^{2}+
s\left(3\,\beta_{\pi}(s)-2\right)}\right]
 -  2\,\mathrm{Li}_{2}\left[\frac{4\, m^{2}-2\, m_{\sigma}^{2}+
s\left(3\,\beta_{\pi}(s)-1\right)}{8\, m^{2}-2\, m_{\sigma}^{2}+
2\, s\left(\beta_{\pi}(s)-1\right)}\right]\nonumber \\
&&\,- \,2\,\mathrm{Li}_{2}\left[-\,\frac{4\, m^{2}-2\, m_{\sigma}^{2}+
s\left(3\,\beta_{\pi}(s)-1\right)}{2\, m_{\sigma}^{2}\beta_{\pi}(s)+\frac{s\,m_{\sigma}^{2}}{2\,m^{2}}
\left(3\,\beta_{\pi}(s)-1\right)\left(\beta_{\pi}(m^{2})-1\right)}\right]
-  2{\,\mathrm{Li}}_{2}\left[\frac{1}{\beta_{\pi}(m_{\sigma}^{2})+\frac{s}{4\,m^{2}}
\left(3\,\beta_{\pi}(s)-1\right)\left(\beta_{\pi}(m_{\sigma}^{2})-1\right)}\right]\nonumber \\
&& \,+\, \mathrm{Li}_{2}\left[\frac{1}{\beta_{\pi}(m_{\sigma}^{2})+\frac{s}{4\,m^{2}}
\left(\beta_{\pi}(s)-1\right)\left(\beta_{\pi}(m_{\sigma}^{2})-1\right)}\right]
 -  2\,\mathrm{Li}_{2}\left[-\frac{4\, m^{2}-2\, m_{\sigma}^{2}+s\left(\beta_{\pi}(s)-1\right)}{2\, m_{\sigma}^{2}\beta_{\pi}(m_{\sigma}^{2})+\frac{s\,m_{\sigma}^{2}}{2\,m^{2}}
\left(\beta_{\pi}(s)-1\right)\left(\beta_{\pi}(m_{\sigma}^{2})-1\right)}\right]\nonumber \\
&& \,+\,\mathrm{Li}_{2}\left[\frac{\beta_{\pi}(s)-3}{\left(\beta_{\pi}(s)-1\right)
\left[\beta_{\pi}(m_{\sigma}^{2})+\frac{s}{4\,m^{2}}\left(3\,\beta_{\pi}(s)-1\right)
\left(\beta_{\pi}(m_{\sigma}^{2})-1\right)\right]}\right]\Biggr\} \,,\\
V_{\pi}(s) & = & -\,\frac{1}{s\,\beta_{\pi}(s)}\Biggl\{ 2{\,\mathrm{Li}}_{2}\left[\frac{-\,2\, m_{\sigma}^{2}+s\left(1-\beta_{\pi}(s)\right)}{s-2\, m_{\sigma}^{2}+s\,\beta_{\pi}(s)\left(\beta_{\sigma}(s)-2\right)}\right]-  2\,\mathrm{Li}_{2}\left[\frac{-\,2\, m_{\sigma}^{2}+s\left(1-3\,\beta_{\pi}(s)\right)}{s-2\, m_{\sigma}^{2}+s\,\beta_{\pi}(s)\left(\beta_{\sigma}(s)-2\right)}\right]\nonumber \\
&& \,+\,2\,\mathrm{Li}_{2}\left[-\frac{2\,m_{\sigma}^{2}+s\left(3\,\beta_{\pi}(s)-1\right)}{s+2\, m_{\sigma}^{2}\beta_{\pi}(m_{\sigma}^{2})-3\, s\,\beta_{\pi}(s)+\frac{s\,m_{\sigma}^{2}}{m^{2}}\left(3\,\beta_{\pi}(s)-1\right)
\left(\beta_{\pi}(m_{\sigma}^{2})+1\right)}\right]\nonumber \\
&& \,-\,2\,{\mathrm{Li}}_{2}\left[\frac{4\, m^{2}-2\, m_{\sigma}^{2}}{s+
2\, m_{\sigma}^{2}\beta_{\pi}(m_{\sigma}^{2})-3\, s\,\beta_{\pi}(s)+\frac{s\,m_{\sigma}^{2}}{2\,m^{2}}\left(3\,\beta_{\pi}(s)-
1\right)\left(\beta_{\pi}(m_{\sigma}^{2})+1\right)}\right]\nonumber \\
&& \,+\,\mathrm{Li}_{2}\left[\frac{\left(4\, m^{2}-2\, m_{\sigma}^{2}\right)s\left(\beta_{\pi}(s)-3\right)}{s\left(\beta_{\pi}(s)-1\right)\left[s+2\, m_{\sigma}^{2}\,\beta_{\pi}(m_{\sigma}^{2})-3\, s\,\beta_{\pi}(s)+\frac{s\, m_{\sigma}^{2}}{2\,m^{2}}\left(3\,\beta_{\pi}(s)-1\right)\left(\beta_{\pi}(m_{\sigma}^{2})+1\right)\right]}\right] \nonumber \\
&& \,+ \,\mathrm{Li}_{2}\left[\frac{4\, m^{2}-2\, m_{\sigma}^{2}}{s+2\, m_{\sigma}^{2}\beta_{\pi}(m_{\sigma}^{2})-s\,\beta_{\pi}(s)+\frac{s\, m_{\sigma}^{2}}{2\,m^{2}}\left(\beta_{\pi}(s)-1\right)\left(\beta_{\pi}(m_{\sigma}^{2})+1\right)}\right]\Biggr\}\,.
\end{eqnarray}
\end{widetext}

\begin{figure*}
\begin{center}
\includegraphics[width=12cm]{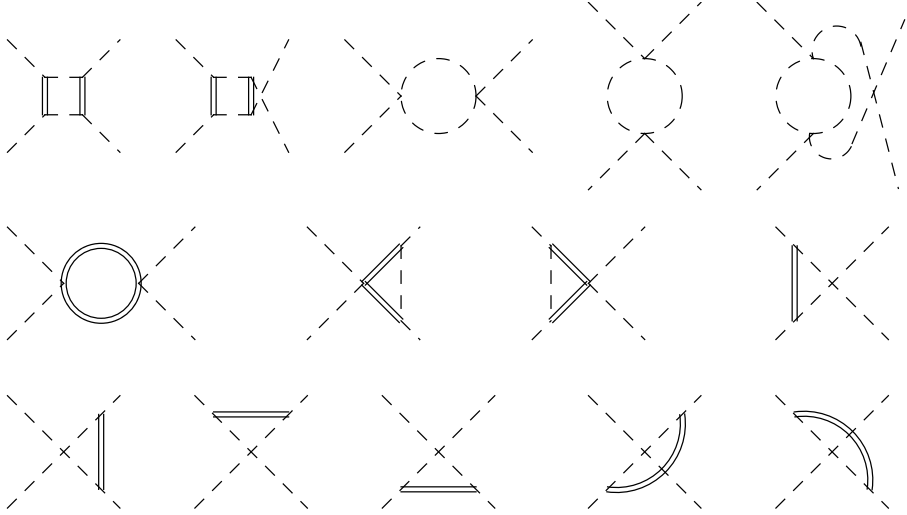}
\end{center}
\caption{Quantum corrections to the four-pion irreducible function.\label{fig:four-point}}
\end{figure*}

The last set of diagrams to be considered are the corrections to the four-pion vertex, that is, the four-point irreducible function. Diagrams are shown in Fig~\ref{fig:four-point} and only contribute to the $\pi\,\pi$ scattering. The structure of the amplitude for the process $\pi^{a}\,\pi^{b}\,\to\,\pi^{c}\,\pi^{d}$ is identical to that of Eq.~(\ref{eq:chew-man}), and our result corresponds to $A(s,t,u)$. The renormalized result is
\begin{eqnarray}
A^{4\pi} & = & -\,2\, g+\frac{g^{2}}{8\,\pi^{2}}
\biggl\{ 2\left(m_{\sigma}^{2}-m^{2}\right)^{2}\left[D(s,t)+D(s,u)\right]\nn
&&+\,
4\left(m_{\sigma}^{2}-m^{2}\right)\left[V_{\pi}(s)+V_{\sigma}(s)\right] +  V_{\sigma}(t)+V_{\sigma}(u)\nn
&&+\,I_{\sigma\sigma}(s)+7\, I_{\pi\,\pi}(s)+
2\left[I_{\pi\,\pi}(t)+I_{\pi\,\pi}(u)\right]\biggr\} \,,\nn
\end{eqnarray}
where $D(s,t)$ is the scalar four-point one-loop function, or scalar box diagram, with all external momenta set to $m^{2}$ and two internal masses equal to $m$ and the other two equal to $m_{\sigma}$, as can be deduced from Fig.~\ref{fig:four-point}. Its expression is rather cumbersome and will not be displayed here, but it can be found, for instance, in Ref.~\cite{Denner:1991qq}.

All the pieces must be combined together to give the one-loop amplitude. First we recall the tree-level amplitude
\begin{equation}
A(s,t)_{\text{tree-level}}\,=\,-\,2\,g\,
\dfrac{s-m^{2}}{s-m_{\sigma}^{2}}\,,
\end{equation}
which reduces to the well-known $\mathcal{O}(p^2)$ $\chi$PT result when the $m_{\sigma}\to\infty$ limit is taken. The renormalized one-loop amplitude is then
\begin{widetext}
\begin{eqnarray}
{A(s,t)}_{1\mathrm{-loop}} & = & \frac{g^{2}}{8\,\pi^{2}}\,
\biggl\{ 2\left(m_{\sigma}^{2}-m^{2}\right)^{2}\left[D(s,t)+D(s,u)\right]+
4\left(m_{\sigma}^{2}-m^{2}\right)\left[V_{\pi}(s)+V_{\sigma}(s)\right]
+  V_{\sigma}(t)+V_{\sigma}(u)\nn
&&+\,I_{\sigma\sigma}(s)+7\, I_{\pi\,\pi}(s)+
2\left[I_{\pi\,\pi}(t)+I_{\pi\,\pi}(u)\right]
 +  \frac{3\left(m_{\sigma}^{2}-m^{2}\right)^{2}}{\left(m_{\sigma}^{2}-
s\right)^{2}}\left[I_{\pi\,\pi}(s)+3\, I_{\sigma\sigma}(s)-3\, A_{\sigma}\right]\nn
&&+\,  2\,\frac{\left(m_{\sigma}^{2}-m^{2}\right)}{s-m_{\sigma}^{2}}\left[2\left(m_{\sigma}^{2}-
m^{2}\right)V_{\sigma}(s)+6\left(m_{\sigma}^{2}-m^{2}\right)V_{\pi}(s)
+  4\, I_{\sigma\pi}(m^{2})+5\, I_{\pi\,\pi}(s)+
3\, I_{\sigma\sigma}(s)\right]\biggr\} \,,\nn
\end{eqnarray}
and the total amplitude to one loop is given by adding the two. It is $\mu$-independent once we take into account the running coupling constant of Eq.~(\ref{eq:running}).
\end{widetext}

\end{appendix}

\end{document}